\documentclass[aps,preprintnumbers,prl,twocolumn,superscriptaddress]{revtex4}

\usepackage{epsfig,latexsym,cancel,amssymb,amsmath}
\usepackage{graphicx}
\usepackage{subfigure} 
\usepackage{epstopdf}
\usepackage{hyperref, graphicx, slashed, xcolor, amsmath}
\usepackage{natbib}
\usepackage{tensor}
\allowdisplaybreaks

\begin{document}

  \title{Gravitational Waves from Magnetars}
  \author{Chris Kouvaris}
\affiliation{Physics Division, National Technical University of Athens, 15780 Zografou Campus, Athens, Greece} 

\begin{abstract}
We study the emission of gravitational waves produced by the magnetosphere of magnetars. We argue that several features in the spectrum could facilitate the identification of that source. In addition,  in cases of extremely large magnetic fields we demonstrate that this emission can contribute to the deviation of the braking index from 3, which is the standard prediction of the magnetic dipole radiation and aligned rotator mechanisms. A similar picture arises if one focuses on the second braking index. Moreover the braking index depends in a specific way on both the rotational frequency and the strength of the magnetic field which could facilitate the identification of this mechanism with respect to others. We also show that gravitational waves can be produced by  polar gap regions due to  their rapid charge-discharge process that takes place in timescales from nanosec to microsec. This can provide an alternative way to probe magnetars and test the polar gap model. 
\end{abstract} 

 \maketitle

Since the first detection of gravitational waves (GW) a few years ago~\cite{LIGOScientific:2016aoc}, a lot of effort has been concentrated on this front both theoretically and experimentally. With the advent of Pulsar Timing Arrays and a series of new interferometers that have been proposed and expected to enter operation in the near or far future, there has been a lot of interest for new astrophysical GW sources that can potentially reveal a lot of information regarding astrophysical and cosmological processes. One field that has been studied extensively for obvious reasons is the  production of GW by neutron stars (NS).  Due to their compactness, there is a lot of opportunity to emit GW under certain conditions.  NS can get involved in binaries among  themselves or with black holes. This is the strongest type of signal  that can come from a NS and this has already been detected~\cite{LIGOScientific:2017vwq}. Other possible cases of GW emission by NS is via several non-radial oscillations modes that can be excited in principle either by internal processes or by encounters with bypassing objects.   A rotating NS is more oblate in the equator. This by itself cannot produce any GW, since the latter requires a time varying quadrupole (or higher) moment. Therefore in case for example a bypassing object disturbs the rotation of the NS, a nonzero second derivative of the quadrupole moment can be created, leading to emission of GW. For gravitational waves produced by pulsating NS, see e.g.~\cite{Andersson:2000mf}.  Another interesting possibility that could lead to an emission of a continuous GW signal is the existence of a thermal and/or magnetic mountain (see \cite{Gittins:2024zbg} for a review).  Thermal mountains
can arise in situations where the NS accretes matter from a companion which compresses underlying layers, triggering nuclear reactions that alter locally the energy density and thus creating a quadrupole moment~\cite{Ushomirsky:2000ax}.  Magnetic mountains can appear and produce GW in cases where there is a misalignment between magnetic and rotational axis.  NS with large toroidal magnetic fields distort the NS into a prolate shape that can cause continuous GW emission~\cite{Cutler:2002nw,Glampedakis:2012qp}.

There is yet another  subtle mechanism where GW can be produced by NS  involving neither some internal process nor a bypassing object. As it is well known since the 60's, a rotating NS with a magnetic field cannot be without the presence of a magnetosphere~\cite{GJ}. In case where the rotation and magnetic axes do not coincide, the co-rotating magnetosphere can create GW that in principle can be detected in future experiments. This scenario has been already studied e.g. in~\cite{Nazari} and~\cite{Contopoulos} where the GW produced by the magnetosphere were deduced. The focus of these two papers was on the detectability of these GW in upcoming interferometers. Continuous GW produced in the context of glitches has been considered in~\cite{Yim:2023nda}.

In this paper we verify that indeed GW produced by the magnetosphere of NS with strong magnetic fields are within reach in upcoming experiments. In addition we show that this is a source of continuous GW with a spectrum consisting of two frequencies, i.e., the frequency of the NS rotation $f_0$ and $2f_0$ with a very specific relative amplitude between the two. Furthermore we focus on another aspect of this scenario. We study how the emission of such GW by the magnetosphere affects  the first and second braking index of the NS. We shall argue that in case of magnetars with extremely high magnetic fields, the braking index can deviate from 3 which is the number expected in standard scenarios where the NS deceleration takes place due to magnetic dipole radiation and/or  via torques of escaping to infinity currents from the poles of the NS. We find that unlike the usual magnetic dipole and aligned rotator models where the braking index is independent of both the rotation frequency and the magnetic field, in high magnetic field magnetars, the braking index depends on both mentioned quantities. Similar conclusions can be drawn for the second braking index. 
We will show that small deviations from the standard values result from NS with fields of at least $10^{15}-10^{16}$ Gauss. Although such  high magnetic fields have not been observed in magnetars so far, this is not unexpected. Arguments based on the virial theorem suggest fields up to $10^{18}$ Gauss for NS with nuclear cores~\cite{ShapiroLai}, while for magnetars with a quark core this value can rise to $10^{20}$ Gauss~\cite{Ferrer:2010wz}. Hybrid stars with quark and nuclear matter and high magnetic fields have been studied in~\cite{Sotani:2014rva}. The effects of extremely high magnetic fields on the QCD equation of state (relevant also for NS) has been recently readdressed~\cite{Fraga:2023cef,Fraga:2023lzn}.

Finally we study an interesting effect within the framework of the polar gap model. The vacuum region above the polar cap discharges via sparks that create a cascade of positron-electron pairs that screen the electric field within a timescale that ranges from $10^{-9}$ to $10^{-5}$ sec. Within the volume of the polar gap, this quick discharge creates time varying fluctuations in the energy density and quadrupole moment that produce GW with frequencies in the ~MHz-GHz range. Given the fact that the frequency range is much larger than that of the rotation, this effect albeit small in amplitude, could in principle provide a tool to test the polar gap model. It is worth mentioning that in this context, GW are emitted even if the rotation and magnetic axes are antiparallel.

\section{Gravitational Waves produced by the Magnetosphere}
As already mentioned the production of GW due to the co-rotating motion of the magnetosphere in the case where rotation and magnetic axes are not aligned has been studied in~\cite{Nazari} and~\cite{Contopoulos}. Here we repeat the calculation within the quadrupole approximation. This introduces a minor as it turns out error in the estimate of the GW compared e.g. to ~\cite{Contopoulos} where the calculation does not implement this approximation. However the quadrupole approximation allows us  to provide results for any angle $\alpha$ between the rotation and magnetic axes and not just only for the  case of $\alpha=\pi/2$. In addition we do not implement a slow rotation approximation  ignoring higher powers of $\omega R$ ($\omega$ and $R$ being the angular frequency and radius of the NS) as in~\cite{Contopoulos}. 

We model the magnetic field of the star to be given by a typical magnetic dipole of the form
\begin{equation}
\vec{B}=\frac{3\hat{r}(\hat{r} \cdot \vec{m})-\vec{m}}{r^3},
\label{magnetic}
\end{equation}
where $\vec{m}$ is the magnetic dipole with magnitude $B_0 R^3$ ($B_0$ being the magnetic field at the surface of the star). We assume that the magnetic axis  subtends an angle $\alpha$ with the rotation axis and therefore $\vec{m}= B_0 R^3 (\sin\alpha \cos\omega t, \sin\alpha \sin\omega t, \cos\alpha)$. Upon the assumption that the NS is to a good approximation a perfect conductor, the presence of the magnetic dipole field in a rotating NS induces an electric field of the form  
\begin{equation}
\vec{E}=-(\vec{\omega} \times \vec{r}) \times \vec{B},
\end{equation}
in order for the Lorentz force applied to every charged particle  to be zero in the co-rotating frame. We use $c=1$. As it was demonstrated in~\cite{GJ} in a rotating NS with a magnetic field, the induced electric field  causes particles to detach from the star and fill up the surrounding space, thus creating a magnetosphere.
The individual components of the electric field are
\begin{equation}
\begin{aligned}
E_r & =  -\frac{m \omega}{2 r^2}(\sin\alpha \sin 2\theta \cos (\omega t -\phi)-2\cos\alpha \sin^2\theta), \\  
E_\theta & = -\frac{2m \omega}{ r^2}\sin\theta (\sin\alpha \sin\theta \cos (\omega t -\phi)+\cos\alpha \cos \theta), \\ 
E_{\phi} & =  0. 
\end{aligned}
\label{electric1}
\end{equation}
The contribution to the  energy density from the electric and magnetic fields is $\rho=(E^2+B^2)/(8\pi)$ where the components of both fields can be read off by Eqs.~(\ref{magnetic}) and (\ref{electric1}). The quadrupole moment is 
\begin{equation}
Q_{ij}=\int d^3x \rho (x,t)(x_i x_j-\frac{1}{3}\delta_{ij} r^2).
\label{quadrupole}
\end{equation}
Since the magnetosphere co-rotates with the star, the integration is limited within the light cylinder i.e., at distances up to $\omega r \sin\theta =1$.
Given that the time dependence of the quadrupole moment enters in the time variation of the energy density, we can calculate both the amplitude of the GW and the energy emitted in that form:
\begin{equation}
h_{ij}^{TT}=\frac{2 G}{r}\Lambda_{ij,kl}\ddot{Q}_{kl}(t-r),
\end{equation}
where $\Lambda_{ij,kl}$ is the appropriate projection operator for the transverse traceless gauge. Using Eqs.~(\ref{magnetic}) and (\ref{electric1}) we can calculate the energy density and subsequently the the second derivative of the quadrupole moments by use of Eq.~(\ref{quadrupole}), getting 
\begin{equation}
\begin{aligned}
\ddot{Q}_{11} & =-\ddot{Q}_{22}=A B_0^2 R^5 \omega^2 \cos 2\omega t \sin^2\alpha, \\
\ddot{Q}_{12} & =A B_0^2 R^5 \omega^2\sin 2\omega t \sin^2\alpha, \\
\ddot{Q}_{13} & = C B_0^2 R^5 \omega^2 \sin 2\alpha \cos\omega t, \\
\ddot{Q}_{23} & = C B_0^2 R^5 \omega^2 \sin 2\alpha \sin\omega t, \\
\ddot{Q}_{33} & = 0.
\end{aligned}
\label{Qs}
\end{equation}
\begin{figure}[tp]
\includegraphics[scale=.4]{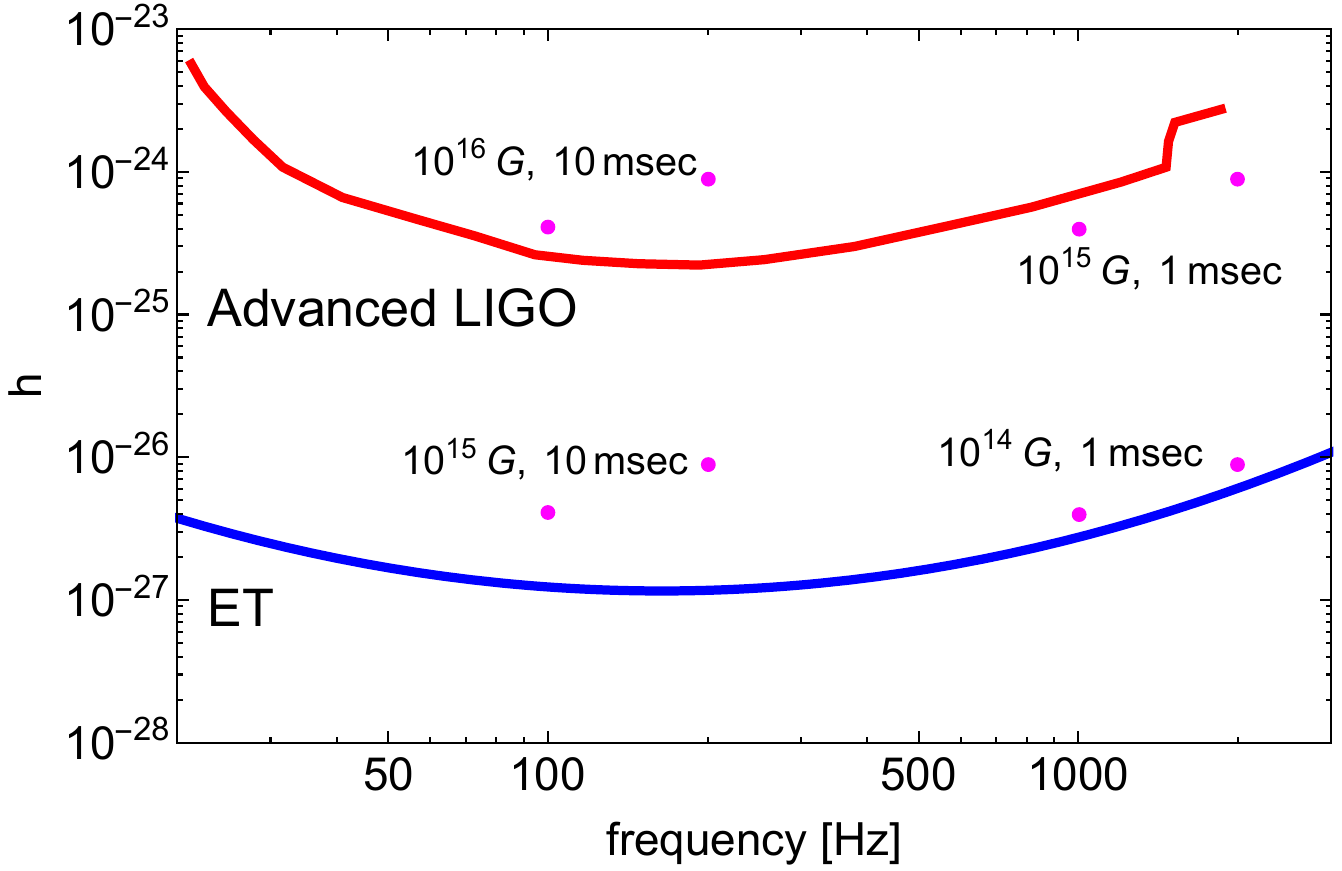}
  \caption{GW amplitude produced by the magnetosphere for NS of various magnetic fields and rotation periods. We have chosen $\alpha=\psi=\pi/4$. Each choice appears as a doublet in the frequency domain, one with frequency that of the NS rotation $f_0$ and 2$f_0$. Note that for $\alpha=\pi/2$ only the $2f_0$ component survives. The signal is calibrated with sources at a distance of 1 kpc.}
  \label{fig:h}
  \end{figure}
  The coefficients $A$ and $C$ are given by
\begin{equation}
\begin{aligned}
A & =-\frac{2}{5}-\frac{3\pi \omega R}{64}+\frac{10 \omega^2 R^2}{21},\\
C &=-\frac{1}{10}+\frac{2 \omega^2 R^2}{35}.
\end{aligned}
\end{equation}
 Note that in the case where $\alpha \neq  \pi/2$ (and $\alpha \neq 0,\pi$), there are two
frequencies emitted i.e., $\omega$ and $2\omega$, contrary to the case where $\alpha=\pi/2$ where only the $2\omega$ is emitted. This is quite important in recognising and identifying such a signal. In the most probable scenario $\alpha \neq \pi/2$ and therefore it is of crucial importance for interferometers to not only detect the two frequencies but also the relative strength between the two. To this end, upon assuming that the detector lies in a direction forming an angle $\psi$ with the rotational axis of the NS and having chosen conveniently a corresponding $\phi =0$, we can estimate the two modes of GW~\cite{Maggiore}
\begin{equation}
\begin{aligned}
h_+ & =\frac{G}{r}(\ddot{Q}_{11}-\cos^2\psi \ddot{Q}_{22}-\sin^2\psi \ddot{Q}_{33}+\sin 2\psi \ddot{Q}_{23})= \\
 & =  \frac{G}{r}B_0^2R^5\omega^2 [A(1+\cos^2\psi)\sin^2\alpha\cos 2\omega t \\
 &+  C\sin 2 \psi \sin 2 \alpha \sin \omega t ],\\ 
h_\times & = \frac{2G}{r}(\cos \psi \ddot{Q}_{12}-\sin \psi \ddot{Q}_{13})= \\
 =& \frac{2G}{r}B_0^2R^5\omega^2(A\cos\psi \sin^2\alpha \sin 2 \omega t -C \sin \psi \sin 2 \alpha \cos \omega t).
\end{aligned}
\label{hsynepi}
\end{equation}
This signal makes a very concrete prediction. Apart from the characteristic signal of $\omega$ and $2\omega$ frequencies, a system of interferometers could in principle determine the angle $\psi$. Consequently since $A$ and $C$ depend only on $\omega$ (which will be accurately determined) and $R$, the independent detection of the two modes $h_+$ and $h_\times$ could also determine both 
the angle $\alpha$ and $R$ (which in any case does not vary a lot for a NS). In principle such a signal can test our knowledge with respect to the NS magnetosphere and give us information regarding also the magnetic field. To get a sense of the strength of the GW, we can write the overall factor in the previous equation in appropriate units
\begin{equation}
\begin{aligned}
&\frac{G}{r}B_0^2R^5\omega^2  \simeq \\
& 3 \times 10^{-24}\left ( \frac{B_0}{10^{15}\text{Gauss}} \right)^2\left (\frac{R}{12~\text{km}} \right )^5\left (\frac{\text{msec}}{P} \right)^2\left (\frac{\text{kpc}}{r} \right ),
\end{aligned}
\end{equation}
where $P$ is the period of the rotation measured in milliseconds and $r$ the distance to the NS in kiloparsecs.
\begin{figure*}[tp]
  \centering
  \subfigure{\includegraphics[scale=.84]{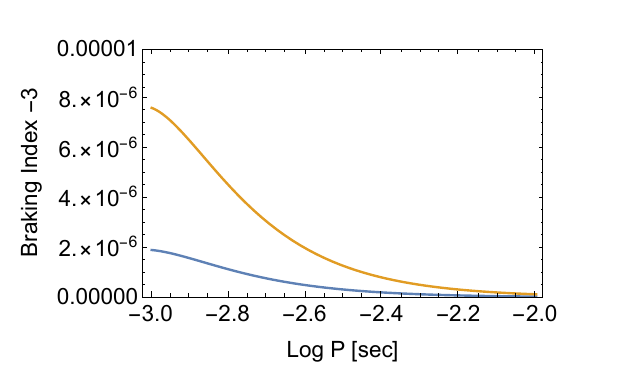}} \quad
  \subfigure{\includegraphics[scale=.83]{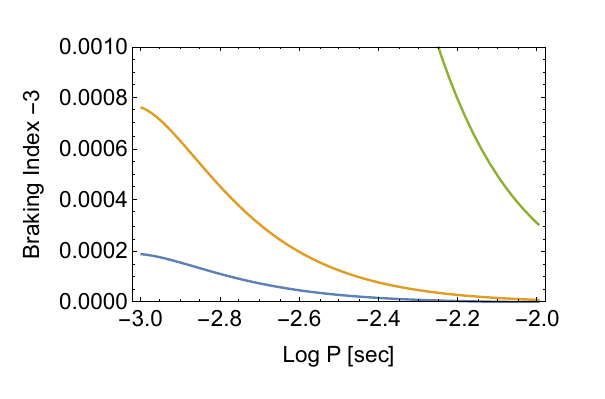}}
  \subfigure{\includegraphics[scale=.83]{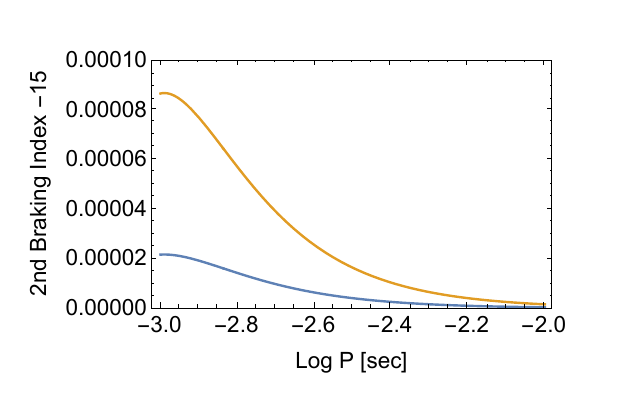}} \quad
  \subfigure{\includegraphics[scale=.83]{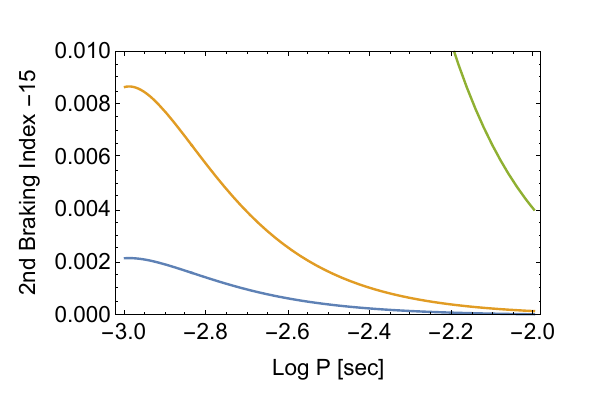}}
  \caption{{\it Upper panels}: Deviation of the first braking index from the standard value 3  for magnetars of different magnetic field as a function of Log P (P being the period of the star in sec). {\it Upper Left panel:} From higher index value to lower, we have stars with
  $B=10^{16}$ and $5\times 10^{15}$ Gauss. {\it Right panel:} The same thing for  $B=5 \times 10^{17}$, $10^{17}$ and $5 \times 10^{16}$ Gauss. {\it Lower panels:} Same as the upper panels  for the deviation of the second braking index from the standard value 15. In all plots we use $R=20$ km and $\alpha=\pi/2$.}
  \label{fig:bi}
\end{figure*}
The amplitude of GW produced by the magnetosphere in the quadrupole approximation is depicted in Fig.~\ref{fig:h} contrasted against the sensitivity of advanced LIGO and the Einstein Telescope. 
Since the signal produced in our studied context is not a burst but rather a continuous wave, we use the advanced LIGO O2 constraints on continuous waves~\cite{LIGOScientific:2019yhl}, where three different methods have been implemented (i.e., FrequencyHough, SkyHough and Time-Domain F-Statistics). Continuous waves are constrained more compared to bursts and inspirals since the observation time can be in principle as long as a year. Similarly we used the  analysis for continuous waves for the Einstein Telescope data~\cite{Branchesi:2023mws}. We converted the constraint on the ellipticity of NS to a constraint on the continuous GW amplitude which consequently turns out to be more accurate (due to the longer observation time) from constraints on inspirals and bursts. For fixed parameters i.e. magnetic field $B$ and frequency of the NS rotation $f_0$, the results depicted in Fig.~\ref{fig:h} come in pairs of two monochromatic GW of frequencies $f_0$ and $2f_0$. One can  see from the figure that e.g. $B=10^{16}$ Gauss NS with periods of $10^{-3}$ sec can already be excluded within distances of kpc.

\section{Braking Index}
NS slow down effectively due to two main mechanisms  i.e., via magnetic dipole radiation and the Goldreich-Julian mechanism of the aligned rotator~\cite{GJ}. In the first case, the fact that the axis of the magnetic field does not align with that of rotation, induces emission of electromagnetic waves via radiation of the precessing magnetic dipole. In the second case, currents of charged particles that are emitted from the caps of the NS in the presence of the magnetic field induce a torque that opposes the motion of the NS, thus acting as an impeding force for the motion  of the star. In general GW can be emitted when there is a time varying quadrupole moment. Although the rotation itself changes the shape of the NS making it oblate, unless there are bumps on the surface of the star, such shape does not create GW because the quadrupole moment remains unchanged in time. However any misalignment between the principal axes of the NS and the rotation axis creates GW. An assessment of possible detection of continuous GW in this case for the Einstein Telescope has been given in~\cite{Branchesi:2023mws}.

However as it has been pointed  out firstly in \cite{Nazari} and \cite{Contopoulos} and we argued in the previous section, the misalignment of the magnetic field with the rotation axis leads to emission of GW that also contribute to the rotational deceleration of the NS even in the absence of misalignment between principal  and rotation axes.

Formally the energy loss of rotational kinetic energy due to emission of electromagnetic waves when both mechanisms are present is~\cite{Contopoulos:2005rs}
\begin{equation}
\frac{dE}{dt}=-L_{\rm mag}\sin^2 \alpha -L_{\rm align}\cos^2\alpha,
\end{equation}
where the terms that appear in the right hand side correspond  to the magnetic dipole and aligned rotator respectively and they are given by
\begin{equation}
L_{\rm mag}=\frac{B_0^2\omega^4 R^6}{4}, \quad \quad L_{\rm align}=\frac{B_0 \omega \omega_{F} R^3 I}{2},
\end{equation}
where $\omega_{F}= \omega -\omega_{\rm death}$, ($\omega_{\rm death}$ being the angular velocity where the pulsar stops  emitting).  The current $I$ has been estimated in the seminal paper of Goldreich and Julian~\cite{GJ} indicating the charge current emitted by the polar caps. It is given by
\begin{equation}
I_{GJ}=\frac{B_0 R^3 \omega^2}{2}.
\end{equation}
Combining the last three equations, taking into account the energy loss due to GW emission and provided that we are interested in angular velocities $\omega >> \omega_{\rm death}$, the overall energy loss reads
\begin{equation}
\frac{dE}{dt}=-\frac{B_0^2 \omega^4 R^6}{4}-G B_0^4 \omega^6 R^{10} f_E(\alpha, \omega R),
\label{dEdt}
\end{equation}
where $f_E$ is an appropriate function of $\alpha$ and $\omega R$. It can be estimated using the fact that the energy loss rate due to GW within the quadrupole approximation is given by
\begin{equation}
\frac{dE}{dt}=-\frac{G}{5}\langle \dddot{Q}_{ij}\dddot{Q}_{ij} \rangle,
\end{equation}
where the brackets denote average over a period.
The first term in the right hand side of Eq.~(\ref{dEdt})  is the sum of the two electromagnetic contributions. Note that since in the limit  $\omega >> \omega_{\rm death}$ their contributions are equal, the $\alpha$ angle dependence drops out. The second term is the contribution coming from the emission of GW. The rate of energy loss is related to the angular velocity deceleration via $dE/dt=-I \omega \dot{\omega}$ where $I$ is the moment of inertia of the rotating NS. The first and second braking indices are defined respectively as 
\begin{equation}
n=\frac{\ddot{\omega}\omega}{\dot{\omega}^2}, \quad \quad m=\frac{\dddot{\omega}\omega^2}{\dot{\omega}^3}.
\end{equation}
Measuring the braking index for a NS is no easy task. It requires the estimate (apart from angular frequency $\omega$ itself) of the first and second derivative of $\omega$. Therefore there is no surprise that 
the braking index has been determined only for a handful of NS (see e.g.~\cite{Espinoza:2011gd,Hamil:2015hqa} and references therein). These measured braking indices are all below 3 and they have been determined with varying  accuracy ranging from 0.001 to 0.1. The second braking index is even harder to measure~~\cite{Parthasarathy:2020ksp}. It requires the estimate of the third derivative of $\omega$ which is known only for the Crab pulsar~\cite{Crab} and PSR B1509-58~\cite{Kaspi}.

For any mechanism where $\dot{\omega}=-k \omega^n$ with $k$ independent of time, the braking index is n and the second braking index is $m=n(2n-1)$. By inspection of Eq.~(\ref{dEdt}) both the aligned rotator model and the magnetic dipole model give $n=3$ and $m=15$. Note also that GW emission provides first and second braking indices $n=5$ and $m=45$ respectively. The co-existence of the electromagnetic emission mechanisms with the GW emission one will make both $n$ and $m$ depart from their corresponding values of 3 and 15. This is depicted in Fig.~(\ref{fig:bi}) where both the first and second braking indices are plotted as a function of the rotational period. As it can be seen larger magnetic fields cause larger deviations from the standard value $n=3$. Starting from $\sim 10^{15}$ Gauss the deviation is $10^{-6}$, culminating to deviations of the order of $10^{-3}$ for fields of the order of $\sim 10^{17}$ Gauss. 
 Such accuracy can be anticipated for some pulsars in the future given the current one which can be as low as $10^{-3}$. As expected, faster rotations cause in general larger deviations from the standard $n=3$ and $m=15$ values. A similar picture is drawn for the second braking index $m$. Note here that for a large part of the studied parameter space, it appears that the deviation from the standard value of $m$ is roughly an order of magnitude larger than that of $n$.
  
 Note also that although in general the deviation for both $n$ and $m$ increases as the period of the rotation decreases, this does happen indefinitely. As we approach msec periods both braking indices do not increase further but rather they start to decrease. This is more pronounced for larger magnetic fields.
  This is in distinct contrast to what happens to the braking index that is dominated by GW emission under the condition that the GW emission is due to misalignment between the principal axes and the rotation one, which is expected to have braking indices monotonically decreasing with the period. This feature is attributed to the nontrivial way that powers of  $\omega R$ enter the expression of the quadrupole moment. It reflects the fact that the GW emission depends on both the rotation and magnetic field. Ordinarily, higher frequency means larger GW amplitude. However in this case, higher frequency leads also to smaller light cylinder and thus smaller volume for the magnetosphere that contributes to the GW emission. 
Another characteristic feature which is distinctly different from the typical GW emission due to the precession  of a thermal mountain  is the fact that in our studied scenario, both indices depend strongly on the magnetic field unlike the former case. Although the distinction of the presented mechanism with respect to GW emission from thermal mountains can be achieved by looking at isolated and non-accreting NS, the situation is more perplexed in the presence of magnetic mountains already mentioned in the introduction. Emission from this mechanism is also continuous in nature, it  depends also on the magnetic field, and can dominate the emission from the magnetosphere. In principle both mechanisms are in place and both contributions should be taken into account. The magnetosphere contribution can be more pronounced  if the NS does not possess large toroidal magnetic fields, condition that minimizes the contribution of the magnetic mountain. Therefore precise knowledge of the NS's magnetic field is important.
The magnetic field of NS is typically estimated by the deceleration rate of the rotation.  However in principle, a different independent estimate of the magnetic field could actually test our scenario. In any case, even if the magnetic field cannot be determined independently, knowledge of both $n$ and $m$ can potentially distinguish this deceleration mechanism from the standard ones (magnetic dipole and aligned rotator) that are independent of the magnetic field since in our studied scenario both $n$ and $m$ depend on the magnetic field nontrivially. It could also potentially distinguish this mechanism from the magnetic mountain one, based on the fact that as the rotational frequency increases beyond a certain point, the former (in contrast to the latter) switches from an increasing GW power to a decreasing one. This is directly related to the aforementioned contraction of the light cylinder that beyond a point results to a reduction in the GW emission of the magnetosphere. The study of different NS with different rotational frequencies and magnetic fields could potentially disentangle the two contributions provided accurate measurements of the toroidal components of the magnetic fields.

As a final comment it should be noted that the inclination angle $\alpha$ and the magnetic fields evolve with time. 
 In such a context the angle $\alpha$ and the magnetic field can be correlated to the rotational frequency. Understanding that correlation could provide another way to test this mechanism since it could make a concrete prediction on braking indices, magnetic field and $\alpha$ simultaneously. However for the moment this is an open problem. Both the time evolution of $\alpha$ and the magnetic field are not well understood basically because we do not know the exact mechanism of magnetic field formation in the star. In particular the time evolution of $\alpha$ can in principle change the values of the braking indices by changing the initial value of $\alpha$ and by contributing an extra term on $\dot{\omega}$ proportional to $\dot{\alpha}$. However the situation is far from clear and has not yet been resolved. Observations of several pulsars~\cite{Weltevrede:2008ic} indicate that there is a correlation between the period (and age) of the pulsar and the angle $\alpha$. Older pulsars seem to be more aligned than younger ones. Note that this study did not include millisecond pulsars which is our target group in this paper. Some 3D simulations of the magnetosphere~\cite{Philippov:2013aha} predict a rapid exponential or milder power law 
 decay of $\alpha$ (depending on the type of the magnetosphere)  taking place in the characteristic slow down timescale $(\dot{\omega}/\omega)^{-1}$. In any case, even for slower rotating pulsars it seems hard to reconcile an exponential or power law decay of the inclination angle with the observed pulsars since significant angles persist at ages where the simulations of~\cite{Philippov:2013aha} predict negligible ones (see e.g. the discussion in~\cite{Johnston:2017wgm}). Even more puzzling is the fact that the Crab pulsar \cite{Lyne:2013voa,Lyne:2014qqa} exhibits an increase  in the alignment angle with a rate of $0.566^0/$year contrary to the expectation of a decreasing $\alpha$ with time. Why $\alpha$ increases for the Crab pulsar and why this happens with that rate is not well established. The inclination angle could increase up to the time where dissipative torques due to viscosity kick in~\cite{Jones}. For that to happen, the star has to cool significantly for $10^4$ or $10^5$ years. The crucial element here is that this timescale is related to the cooling process and not to the slowing down of the star. 
As it is clear from the above discussion, the time evolution of the inclination angle is far from being understood. The purpose of this paper here is not to speculate on these open issues, but rather to demonstrate that under general and robust conditions, the braking index due to GW emission can deviate significantly from the predicted values of the standard scenarios of the magnetic dipole radiation and the aligned rotator. 
Fig. 2 provides braking indices for an angle $\alpha=\pi/2$. Because of the fact that the $\alpha$ dependence of the function $f_E$ enters as powers of $\sin 2\alpha$ and $\sin 4 \alpha$, the derivative of $\alpha$ with respect to time is zero at that exact value and therefore does not contribute an extra term in the braking index. Therefore in the specific case  the results are exact (i.e. the effect of a nonzero $\dot{\alpha}$ is taken into account). As a further cross-check the braking index for a different value e.g., $\alpha=\pi/4$ has been calculated, using a rate of change for $\alpha$ similar to the observed one from the Crab pulsar which has several of the desired properties (large magnetic field $\sim10^{13}$ Gauss field and a small 33 msec period). Including the effect of the time evolution of inclination angle is negligible (changes of the order of $10^{-16}$). In fact this does not come as a surprise. The functional dependence of $f_E$ is power law for both $\omega R$ and $\sin \alpha$. Therefore the relative strength of the extra term that involves $\dot{\alpha}$ comes down to the comparison between the timescales $(\dot{\alpha}/\alpha )^{-1}$ and 
$(\dot{\omega}/\omega )^{-1}$. As long as the timescale for the decay of $\alpha$ is much longer than that of the rotational deceleration, the effect will be negligible. Obviously if the timescales become comparable, the contributions become comparable too. However even in that extreme case, it is highly unlike to  have the two contributions fine tuned in a way that one cancels the other. Therefore one  anticipates that the values presented in Fig.~2 even in that special case give the correct ballpark of the braking index. The purpose of this work is to point out that GW emission from the magnetosphere can cause observational changes in the braking index of the magnetars with a nontrivial dependence on the magnetic field in striking contrast to the aligned rotator and the magnetic dipole models as well as the emission of GW due to oblateness. As long as  $(\dot{\alpha}/\alpha )^{-1}>>(\dot{\omega}/\omega )^{-1}$ we can ignore the time dependence of $\alpha$ since it is negligible. We leave for future work the interesting case where the two timescale are comparable.

\section{Gravitational Waves from the Polar Cap}
Due to symmetry and reflected in Eq.~(\ref{Qs}),  when the magnetic field is parallel or antiparallel to  the rotation axis, there is no emission of GW. However even in this case there is a potential source for GW production at least within models that exhibit polar gaps. The aforementioned picture of Goldreich and Julian portrayed earlier where the  rotation of the NS causes emission of particles from the poles is not accurate due to the inability of extraction of  positive charges. As a result, polar gaps might form~\cite{Ruderman}, which inside them there is no charge. This lack of the appropriate Goldreich-Julian charge density allows the development and growth of electric fields that are parallel to the magnetic dipole field. As time goes on, the electric field and the gap increase. It had been shown that the gap cannot increase indefinitely. In fact the maximum height that the gap can reach cannot exceed the size of the polar region with a radius $r_p$ which is basically the radius of a disk centered with the magnetic and rotation axis where magnetic field lines  departing from inside the disk cross the light cylinder. As it was argued in~\cite{Ruderman},  under reasonable conditions, curvature photons that are produced by particles accelerated within the gap can lead to a spark that discharges the gap, i.e., a cascade of positron-electron pair production. The electric field  accelerates positrons and electrons in opposite directions, thus discharging the gap. Ruderman and Sutherland argue that the discharge process takes place in timescales of microsec  which is roughly the time it takes for a relativistic particle to travel across the polar gap multiplied by a factor $\sim 40$. Once discharged and filled with the appropriate Goldreich-Julian  charge density, the electric field drops to zero. The Goldreich-Julian current $I_{\rm GJ}$ quickly depletes once again the charge from the polar gap and the electric field starts growing again. This  happens within a similar timescale since charges are flying off the region with the speed of light. This cycle of charging and discharging the polar cap region in short timescales with rapid changes of extremely high electric fields can potentially induce GW as mentioned earlier even in the case where the star has the magnetic dipole and rotation axis in an antiparallel fashion.

Estimating accurately the GW spectrum is complicated. As the gap grows in height and strength of the electric field, it gets the same time depleted by charge. These charged particles are dispersed in the space of the magnetosphere outside the polar gap. Therefore a precise determination of the GW signal would require two ingredients: i) knowledge of how charged particles are dispersed outside the polar gap as the latter is evacuated and ii) an accurate knowledge of the evolution of the electric field as a function of time and space. Here we make an approximate analysis for the amplitude of the GW produced based on a simplified toy model. Nevertheless, a lot of insight can be gained from such an attempt. On the one hand this studied effect produces GW even in the case where there is no misalignment between rotation and magnetic axes. On the other hand, due to the fact that the timescale involved  in this process is not the period of the NS but the timescale of discharge which is considerably smaller, the GW frequency is of the order of at least $10^5$ Hz. This is a signal at high frequencies, away from the frequency  band of several  GW astrophysical sources. To this end and for the sake of simplicity, for the estimate of the quadrupole moment that will drive the GW production we will consider only the variation of the electric field. The magnetic field  is expected to stay constant and therefore does not source any GW in this scenario. Since we are interested in a ballpark estimate of the signal, we do not consider the complicated change of the density of the charged particles. This will hopefully introduce an error that can be of order one. 

\begin{figure} [htbp]
 \includegraphics[width=1\linewidth]{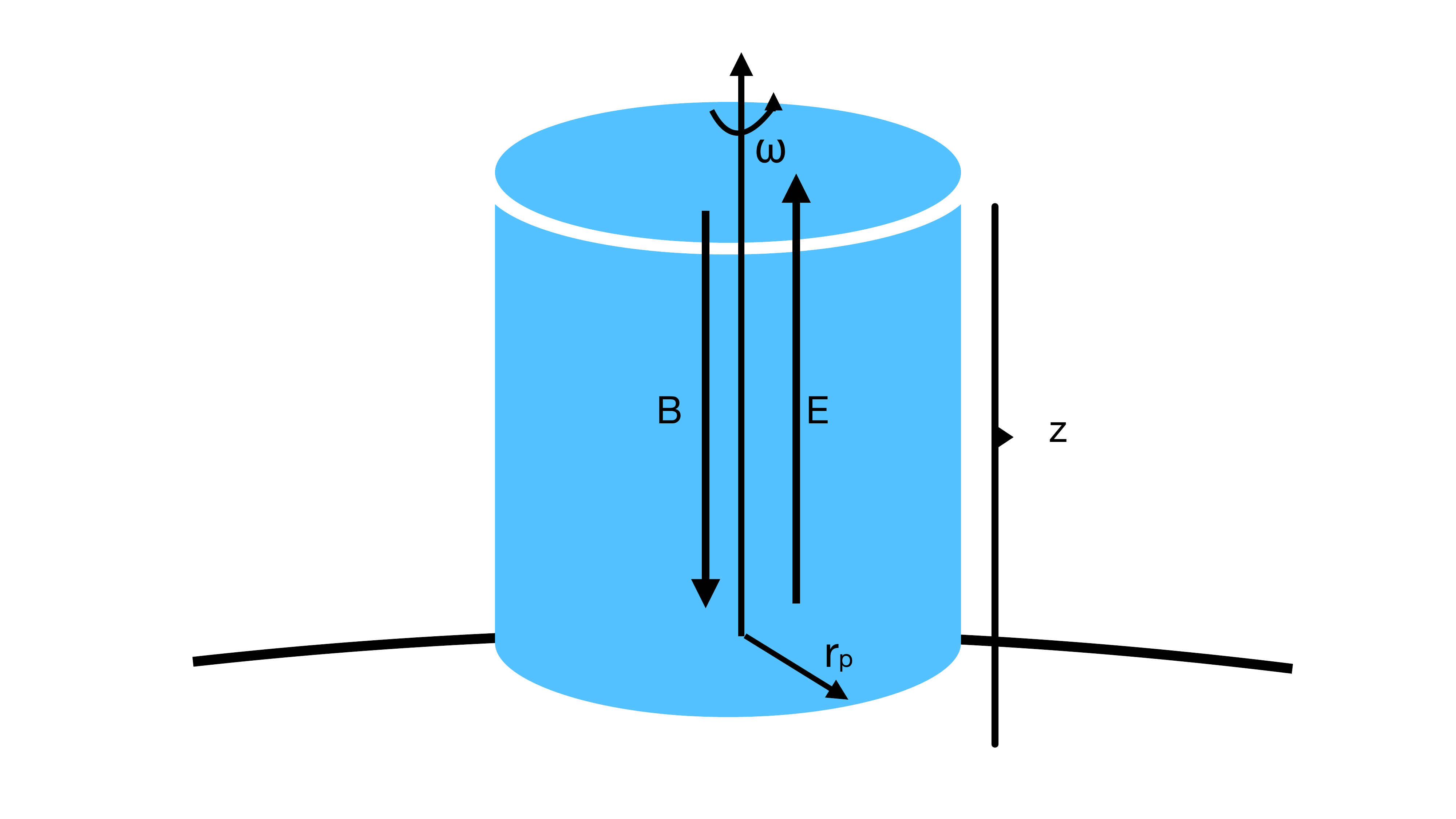}  
    \caption{The polar gap in the case of antiparallel magnetic field and rotational axis. The height of the gap grows linearly with time with the speed of light up to a maximum value $h$, upon which a spark discharges the gap reducing its height and electric field linearly with time until it gets zero.}
\label{fig:pg}
\end{figure}

We adopt a simple model for the variation of the electric field similar to~\cite{Prabhu:2021zve}. We assume that the electric field evolves linearly with time and that the gap  grows also linearly in height with the speed of light up to a maximum height $h$. This is justified by the fact that the Goldreich-Julian current evacuates charges practically with the speed of  light. This depletion of charge is what creates the electric field component parallel to the $z-$axis. Within our approximation the electric field is
\begin{equation}
 E(z)=E_{\rm max}\frac{z-R}{h},  
\label{E2}
\end{equation}
where $z$ denotes the coordinate along the rotation axis. Note that during the charging of the gap $t=z-R-T/2$ for $-T/2\leq t\leq -T/2+h$. $T$ is the total period of the cycle charge-discharge that we will estimate later on. Within our approximation after the electric field reaches its maximum, it will stay there up to time $t=T/2-h$ where the spark ignites. After that moment the charge redistributes and reaches the Goldreich-Julian value that screens completely the electric field. Therefore after $t=T/2-h$ and until $T=T/2$, we have $t=T/2+R-z$.
Note also that $z$ is measured from the center of mass which in good approximation is taken to be the center of the star.
Fig.~(\ref{fig:pg}) illustrates the geometry of the polar cap region. 

Within our setup we can estimate the quadrupole moment of the polar gap. For example
\begin{equation}
M_{11}=\int \frac{1}{8\pi} E(z)^2 x^2 dxdydz,
\end{equation}
where the integration takes place within the cylinder of radius $r_p$ and height $z$. Here $M_{ij}=\int d^3x \rho (x,t) x_i x_j$ is the quadrupole moment without the second term inside the integral of Eq.~(\ref{quadrupole}) that makes it traceless. 
Given the dependence on $z$ of the electric field (see Eq.~(\ref{E2})), we have 
\begin{equation}
\begin{aligned}
M_{11} & =M_{22} \simeq \frac{E_{\rm max}^2 r_p^4}{96 h^2} f(t)\\
M_{33} & \simeq   \frac{E_{\rm max}^2 r_p^2 R^2}{24 h^2}  f(t),\\
\end{aligned}
\label{M1133}
\end{equation}
with 
\begin{equation}
f(t)= 
    \begin{cases}
      \left ( t + \frac{T}{2} \right )^3 & -\frac{T}{2} \leq t \leq -\frac{T}{2}+h \\
      h^3  &-  \frac{T}{2}+h \leq t \leq \frac{T}{2}-h \\
       \left (\frac{T}{2}-t \right )^3 & \frac{T}{2}-h \leq t \leq \frac{T}{2} \\
              \end{cases}.
\end{equation}
Since the function $f(t)$ is periodic, one can expand it in Fourier series of frequencies $n f_0$ where $f_0=1/T$. After that, the second time derivative  can be taken in order to estimate the amplitude of the GW.  However, it is not expected that the process of  charge/discharge is strictly periodic. This deviation from strict periodicity will smooth out the spectrum. In addition one could only take the derived spectrum as a very rough estimate given the crudeness and the uncertainty of the approximation we used to model the whole  charge-discharge cycle. In all the benchmark points we present (see Table~\ref{table:1}), we find that $T>>h$. Although the fundamental frequency will be $f_0$, by inspection of the function $f(t)$, it is understood that the rapid changes which will induce the largest contribution to the GW production is rather $f_1=1/h$ which is much higher than $f_0$. In fact the Fourier analysis verifies exactly this: the signal appears at $f_0$ but the amplitude peaks at $f_1$. Higher frequencies than $f_1$ can also be unsuppressed. Due to the ambiguity regarding the precise form of $f(t)$ we will give an approximate estimate of the produced amplitude of GW as follows. Taking twice the time derivative of Eq.~(\ref{M1133})  yields
\begin{equation}
\begin{aligned}
\ddot{M}_{11} & =\ddot{M}_{22} \simeq \frac{E_{\rm max}^2}{16}\frac{r_p^4}{h} g(t)\\
\ddot{M}_{33} & \simeq   \frac{E_{\rm max}^2}{4}\frac{r_p^2 R^2}{h}  g(t)\\
\ddot{M}_{12} &=\ddot{M}_{13}=\ddot{M}_{23}=0,
\end{aligned}
\label{Mdot}
\end{equation}
where in $\ddot{M}_{33}$ we have omitted terms of the order of $h/R$ and $T/R$. The function $g(t)$ within our approximation is given by
\begin{equation}
g(t)=
\begin{cases}
\frac{t+\frac{T}{2}}{h} & -\frac{T}{2} \leq t \leq -\frac{T}{2}+h \\
0 &     -\frac{T}{2}+h\leq t \leq \frac{T}{2}-h \\
     \frac{ \frac{T}{2}-t}{h} & \frac{T}{2}-h \leq t\leq \frac{T}{2}
\end{cases}.
\end{equation}
By inspection of Eq.~(\ref{Mdot}), for relatively slow rotations $\ddot{M}_{33}>>\ddot{M}_{11}$ because $R>>r_p \simeq R\sqrt{\omega R}$. To give an approximate estimate for the GW amplitude, we note that $g(t)$ is of order one for time between $-T/2 \leq t \leq -T/2+h$ and $ T/2-h \leq t\leq T/2$. As we mentioned earlier the Fourier analysis gives a peak at $f_1=h^{-1}$ and a smaller amplitude at $f_0=T^{-1}$, consistent with our argument here. 

In order to determine the ballpark frequencies $f_0$ and $f_1$ of the produced GW we need an estimate of $h$ and $T$. In the standard picture of~\cite{Ruderman}, a gap where the Goldreich-Julian charge density is absent develops, leading to the creation of a potential difference $\delta V=\omega B h^2$. A single particle accelerates within the gap producing a number of curvature photons which each of them in turn creates a pair of positron-electron. The mean free path for these curvature photons is exponentially sensitive to the magnetic field and the size of the gap $h$. Due to that, once the mean free path becomes $\sim h$ it will lead to an exponential production of pairs that will ignite the spark that will discharge the gap. This happens when~\cite{Ruderman}
\begin{equation}
h\simeq 5 \times 10^3 \rho_6^{2/7}P^{3/7} B_{12}^{-4/7}~\text{cm},
\label{h2}
\end{equation}
where $\rho_6$ is the magnetic field curvature length measured in units of $10^6~\text{cm}$, $B_{12}=B/(10^{12}~\text{Gauss})$ and $P$ the period of the NS rotation in sec. The charged particles accelerate within the gap to energies
\begin{equation}
E_{\rm gap} \simeq \omega B h^2=1.6 \times 10^{12} B_{12}^{-1/7}P^{-1/7}\rho_6^{4/7}~\text{eV}.
\end{equation} 
This energy leads to a relativistic factor $\gamma = E_{\rm gap}/m_e$ where $m_e$ is the mass of the electron. Each of these charged positrons and electrons will create along their path inside the gap an average number of curvature photons
\begin{equation}
N_{\gamma}\simeq \frac{4}{9}e^2\frac{h}{\rho}\gamma.
\label{Ng}
\end{equation}
We can give now a rough estimate of the timescale needed to produce enough charge in the gap vacuum to discharge it. Within the timescale of $h$ which is the time it takes for a particle to cross the gap, $N_{\gamma}$ curvature photons are produced, which at time $2h$ each of them has produced one electron-positron pair. The amount of required charge in order to discharge the gap is given by the Goldreich-Julian density which over the poles is $ \rho_e=-\vec{\omega}\cdot \vec{B}/(2\pi)$.   Therefore the amount of charge needed is $|Q|=\rho_e \pi r_p^2 h$. Within this simple argument, the number of pairs grows as a geometrical series every $2h$, leading to $T\sim 2h\ln [(2N_{\gamma}-1)Q/(2N_{\gamma}e)]/\ln(2N_{\gamma})$. For example for a NS with $B_{12}=P=\rho_6=1$, we get $T\sim 24 h$, while 
for $B_{12}=\rho_6=1$ and $P=10^{-2}$, we get $T\sim 38$. These results are comparable to the range given in~\cite{Ruderman}. However we should stress that in order to discharge the gap, two conditions have to be fulfilled: i) the mean free path for converting the curvature photons to electron-positron pairs has to be smaller than $h$ and ii) $N_{\gamma}$ must be larger than 0.5 or more in order for the pair creation to grow at each time $h$. $h$ cannot become larger than $r_p$ and therefore as a NS slows down and $\omega$ decreases, the gap increases. When the gap becomes $h\sim r_p$, the pulsar ceases to emit because there can be no discharge and production of electron-positron pairs. We are interested in NS with large magnetic fields. As the magnetic field increases,  Eq.~(\ref{h2}) shows 
that $h$ drops and subsequently $N_{\gamma}$ drops too (see Eq.~(\ref{Ng})). Therefore at some magnetic field, $N_{\gamma}$ becomes smaller than 0.5 and there is no cascade of electron-positron pairs created. In this case it is expected that $h$ will increase as much needed, as long as $h\leq r_p$, until $N_{\gamma}$ becomes of order 1.

\begin{table}[t]
\begin{tabular}{ |p{1.8 cm}|p{1.3 cm}|p{1.5cm}|p{1.5cm}|p{1.8cm}|  } 
 \hline
  $B~(\text{Gauss})$   & $\rho_6$ & $f_0~(\text{MHz})$ & $f_1~(\text{GHz})$  & $h_+$    \\
 \hline
 $10^{13}$  &  1  & 3.2   & 0.4 &  $1.4 \times 10^{-32} $   \\
  $10^{14}$    &  0.01     & 0.9 & 6 & $1.02 \times 10^{-31}$  \\
 $10^{15}$  &  1  & 11.1   & 2.2 &  $2.7 \times 10^{-29}$  \\
 $10^{16}$  &  1  & 23.6   & 4.8 &  $1.2 \times 10^{-27}$  \\
  \hline
\end{tabular}
\caption{Benchmark points for different magnetic fields and curvature parameter $\rho_6$. $f_0$ is the fundamental frequency given by the inverse timescale needed for the charge/discharge cycle $T^{-1}$ and $f_1$ is the frequency corresponding to the rapid timescale $h^{-1}$ where the signal becomes maximum (see text). $h_+$ is the strength of the GW. For all cases the rotation period of the NS has been taken to be $10^{-3}$ sec and the distance to the source 1 kpc.
}
\label{table:1}
\end{table}

  The $h_+$ and $h_{\times}$ can be estimated by use of Eq.~(\ref{hsynepi}). Note that this equation holds also if we replace $Q_{ij}$ with $M_{ij}$. Because $\ddot{M}_{12}=\ddot{M}_{13}=0$, we have $h_{\times}=0$. Therefore a very characteristic feature of this emission mechanism (for $\alpha=\pi $) is the absence of $h_{\times}$ modes. The $h_+$ mode is given by
\begin{equation}
h_+=\frac{G}{r}(\ddot{M}_{11}-\ddot{M}_{33})\sin^2\psi.
\end{equation}
As we mentioned $\ddot{M}_{11}<<\ddot{M}_{33}$.  The above equation gives
\begin{equation}
h_+=2 \times 10^{-29} P_3^{-3}B_{15}^2h_{10} r_1^{-1} \sin^2\psi,
\end{equation}
where $P_3$ is the period of rotation in units of $10^{-3}$ sec, $B_{15}$ the magnetic field on the poles in units of $10^{15}$ Gauss, $h_{10}$ the gap in units of 10 cm and $r_1$ the distance to the NS in kpc. For this formula we have assumed $R=20$ km and $\rho_6=1$. Recall that $h$ is not a free parameter but it is the maximum between the value of Eq.~(\ref{h2}) and the minimum  value required in order to have $N_{\gamma}\sim 1$, so the discharge process can take place.  
Ignoring the factor $\sin^2\psi$ and setting the distance to the source to 1kpc, we can estimate the amplitude for different benchmark points presented in Table~\ref{table:1}. $f_0$ and $f_1$ are the two frequencies mentioned earlier. The amplitude for $h_+$ that is presented refers to $f_1$. Depending on the exact form of the functions $f(t)$ or $g(t)$, the amplitude at $f_0$ will be somewhat suppressed compared to that at $f_1$ that we quote at the table. For all benchmark points we used a period for the magnetar of $10^{-3}$ sec. Although for the moment there are no current experiments probing the MHz-GHz range of frequencies, several proposals  for future experiments  exist, aiming exactly at detection of high frequency GW~\cite{Aggarwal:2020olq}. Table 1 of~\cite{Aggarwal:2020olq} gives a summary of all the high GW frequency proposals. Among the different proposals, detectors based on enhanced magnetic conversion could in principle operate at $\sim 10$ GHz with a sensitivity for the strain $10^{-30}$ to $10^{-26}$. Similarly for superconducting rings, the projected sensitivity at 10 GHz is $10^{-31}$. By inspection of Table~\ref{table:1} magnetic fields as low as $10^{14}$ Gauss are above the threshold in the GHz frequency range of the aforementioned proposals. 

We can now generalize the emission of GW for an angle that is not $\alpha=\pi$. If we call the coordinates we used in the problem which we studied above where the rotation axis is antiparallel to the magnetic field axis $x_i'$, we can relate them to the fixed space coordinates $x_i$ where generically the magnetic field subtends an angle $\alpha$ with the rotation axis via $x_i'=R_{ij}x_j$, where $R=R_x{(\pi-\alpha)}R_z(\omega t)$ is a rotation transformation matrix which is the product of a rotation around $x$-axis by angle $\pi -\alpha$ times a rotation around $z$-axis by $\omega t$ ($\omega$ being the angular rotational frequency of the NS).  The quadrupole tensor in the two coordinates systems is related via $Q=R^{\intercal}Q'R$. Once the quadrupole tensor is known, one can estimate the GW spectrum in the generic case.  There are two main observations regarding the generic case. The first one is that for $\alpha \neq \pi$, $h_\times$ is no longer  zero. The second one is that apart from the frequency $\Omega =2\pi/T$ (which is related to the charge-discharge process), there are also the frequencies $\Omega \pm \omega$ and $\Omega \pm 2\omega$. In general $\Omega>>\omega$ and therefore all frequencies lie close to each other. However the combination of all these frequencies makes a very characteristic signal which if and once the high frequency (MHz-GHz) range of GW is probed, it could facilitate the identification of this polar gap effect.

In this paper we studied the GW that can be induced by the magnetosphere of rapidly rotating magnetars. We found that this is within reach in upcoming detectors under sensible assumptions. We also found that this GW emission  can affect the first and second braking index of magnetars as long as the magnetic fields are strong.  Finally we studied the effect of the charging and discharging in the polar gap model. This process can lead to production of GW with frequencies in the MHz and GHz range which currently is not probed by any experiment, although several future proposals and upcoming detectors are in play. The detection of such a signal will be ale to test the polar gap models especially on magnetars.

 Several directions can be taken in the future in order to improve our estimates. One in principle should attempt to include in the changing quadrupole moment of the polar gap, the incoming and outgoing plasma contribution which  could change the amplitude of the GW signal by a factor of order one. This is not a priori an easy task since it would require probably some magnetohydrodynamics simulations of the plasma. Furthermore a more accurate description of the charge-discharge process should be implemented. This has been studied to some extent in~\cite{Timokhin:2010fe,Timokhin:2012sk,Tolman:2022unu}.
For example sophisticated simulations of the process where curvature photons decay to electron-positron pairs which in turn accelerate within the polar gap and emit new curvature photons has been considered in~\cite{Noordhuis:2022ljw}. An implementation of a more accurate time evolution of the electric field could significantly change the shape of the power spectrum of the emitted GW but it will not probably change significantly the main frequency $f_1$ and the amplitude of the GW. Nevertheless it can be crucial in identifying accurately the effect in future high frequency GW detectors.

\end{document}